\documentclass[journal]{IEEEtran}
\usepackage{cite}
\usepackage[Gray,amssymb]{SIunits}
\usepackage{color}
\usepackage{acronym}
\usepackage{esvect}
\usepackage[hidelinks]{hyperref}
\usepackage{amsfonts,amsmath,amssymb}
\usepackage{dsfont}
\usepackage{nicefrac}
\usepackage{lipsum}
\ifCLASSINFOpdf
  \usepackage[pdftex]{graphicx}
  \DeclareGraphicsExtensions{.pdf,.jpeg,.png}
\else
  \usepackage[dvips]{graphicx}
  \DeclareGraphicsExtensions{.eps}
\fi

\DeclareGraphicsExtensions{.pdf,.jpeg,.png,.jpg}

\acrodef{ev}[{EV}]{electric vehicle}
\acrodef{phev}[{PHEV}]{plug-in hybrid electric vehicle}
\acrodef{bev}[{BEV}]{battery electric vehicle}
\acrodef{v2g}[{V2G}]{vehicle-to-grid}
\acrodef{soc}[{SoC}]{state of charge}
\acrodef{ocv}[{OCV}]{open circuit voltage}
\acrodef{cs}[{CS}]{charging station}
\acrodef{pcc}[{PCC}]{point of common coupling}

\acrodef{nmc}[{NMC}]{Nickel Manganese Cobalt Oxide}
\acrodef{nca}[{NCA}]{Nickel Cobalt Aluminum Oxide}
\acrodef{lco}[{LCO}]{Lithium Cobalt Oxide}
\acrodef{lmo}[{LMO}]{Lithium Manganese Oxide}
\acrodef{lfp}[{LFP}]{Lithium Iron Phosphate}

\acrodef{cccv}[{CCCV}]{Constant Current Constant Voltage}
\acrodef{cpcv}[{CPCV}]{Constant Power Constant Voltage}

\acrodef{erlm}[{ERLM}]{exponential recovery load model}
\acrodef{oclm}[{OCLM}]{oscillatory component load model}
\acrodef{vf}[{VF}]{vector fitting}
\acrodef{vflm}[{VFLM}]{vector fitting-based load model}
\acrodef{oltc}[{OLTC}]{on load tap changer}
\acrodef{mae}[{MAE}]{mean absolute error}
\acrodef{rmse}[{RMSE}]{root mean square error}

\newcommand{\PreserveBackslash}[1]{\let\temp=\\#1\let\\=\temp}

\newcommand{\InRefFig}[1]{Figure~\ref{#1}}

\newcommand{\RefFig}[1]{Fig.~\ref{#1}}

\newcommand{\RefEq}[1]{equation~(\ref{#1})}

\usepackage{array}
\newcolumntype{x}[1]{>{\centering\arraybackslash\hspace{0pt}}p{#1}}
\usepackage{multirow}
\usepackage{enumitem}
\usepackage{stfloats}
\begin{document}
\title{Definition of Static and Dynamic Load Models for Grid Studies of Electric Vehicles Connected to Fast Charging Stations}

\author{%
  Davide del Giudice,~\IEEEmembership{IEEE Member},
  Angelo Maurizio Brambilla,~\IEEEmembership{IEEE Member},
  Federico Bizzarri,~\IEEEmembership{IEEE Senior Member},
  Daniele Linaro,~\IEEEmembership{IEEE Member},
  Samuele Grillo,~\IEEEmembership{IEEE Senior Member} \\
  \vspace{-7mm}%
  \thanks{D.~del Giudice, S.~Grillo, F.~Bizzarri, D.~Linaro, and A.M.~Brambilla are with Politecnico di Milano, DEIB, p.zza Leonardo
    da Vinci, n. 32, 20133 Milano, Italy.  (e-mails:
    \{name.surname\}@polimi.it).}%
   \thanks{Italian MIUR project PRIN~{2017K4JZEE\_006} funded
   	the work of S.~Grillo (partially) and D.~del Giudice (totally).}
}

\maketitle

\begin{abstract}
The growing deployment of electric mobility calls for power system analyses to investigate to what extent the simultaneous charging of \aclp{ev} leads to degraded network operation and to validate the efficiency of countermeasures. To reduce complexity and \textsc{cpu} time, a common approach while performing these analyses consists in replacing \aclp{ev} and their \aclp{cs} with constant \textsc{pq} loads. However, this approach is inaccurate, as the power absorbed by these elements actually depends not only on voltage but also on the state of charge, charging method, cathode chemistry of the battery pack, and converter controls in the \acl{ev} and \acl{cs}.

By considering all these aspects, this article develops a novel static load model and a vector fitting-based dynamic load model for \aclp{ev} connected to fast \aclp{cs}.   These computationally efficient representations can replace the standard constant \textsc{pq} load model of \aclp{ev} to assess more accurately their impact in static and dynamic grid studies. Simulation results of the \textsc{ieee14} system modified by adding fleets of \aclp{ev} prove the accuracy of the proposed models and highlight the shortcomings of the standard \acl{ev} representation as a constant \textsc{pq} load in some cases.
\end{abstract}


\section{Introduction}
\label{sec:intro}

\IEEEPARstart{E}{lectric} mobility is growing at a staggering pace.
In 2021, the global fleet of \acfp{ev}
 exceeded 16.5 million, which
was three times the amount in 2018.
This trend is envisaged to intensify,
culminating in the sale of around 200 million \acp{ev} by 2030 \cite{GEO:2022}.
However, despite reducing
greenhouse gas emissions,
the rising popularity of electric mobility challenges modern
power systems.
Indeed,
unless properly managed, adverse effects of massive \ac{ev} deployment include
larger voltage drops and losses, grid overloads, and a decrease in power quality
\cite{Habib:2018,Nour:2020}.

A vast literature has been produced to analyze these phenomena and consequently develop mitigating measures that allow
postponing or at least minimizing the need for grid reinforcement operations \cite{Arias:2020}.
A popular solution consists in
implementing delayed or controlled charging strategies \cite{Pavol:2019}.
These solutions are usually validated through power flow analyses based on a probabilistic
approach.
For example, in the case of \textit{mobility-aware} charging strategies,  probability density functions are used to describe \acp{ev} charging time, location, duration, as well as
their \ac{soc} before charging \cite{Tehrani:2015,Arobinda:2015}.
While these aspects have been studied at considerable length, relatively less
attention has been paid to the derivation of an equivalent \ac{ev} load model, which is just as essential
to accurately assess the impact of electric mobility on the grid.

In principle, for this purpose one might resort to detailed \ac{ev} models that allow analysing electrical variables at the converter level. However, this would lead to prohibitive \textsc{cpu} times
and a degree of accuracy that is typically excessive for large-scale power system simulations. On the opposite side of the complexity spectrum,
most of the previously mentioned studies reduce \acp{ev}
to mere constant active and reactive power (\textsc{pq}) loads \cite{Wang:2020}. This simplification
is in line with \cite{Milanovic:2013}, according to which about 70\% of utilities and system operators
worldwide rely on \textsc{pq} load models for steady-state power system studies.
Such a modelling approach
has become common
practice because it is easy to implement, has a low computational burden, and generally
implies considering a worst-case scenario, thereby providing results that are well within safety boundaries \cite{delGiudice:2022,Kontis:2022}.

Nonetheless, there are some
exceptions arguing that the constant \textsc{pq} load
model does not adequately reflect the real behaviour of \acp{ev}, not even at steady-state.
In the context of static load modelling, the works in
\cite{Haidar:2016} and \cite{Aguirre:2019} describe \acp{ev} with a \textsc{zip} model by adopting a
measurement-based approach.
Other works resort to a \textit{multistage}
\textsc{zip} model,
where piecewise functions describe the \ac{ev} active and reactive power exchange;
each interval is defined by different \textsc{zip} model parameters and
represents a specific \ac{ev} working condition (e.g., charging time \cite{Haidar:2015}
or \ac{soc} \cite{Sortomme:2012,Sortomme:2013}). Lastly,
\cite{Saha:2014} employs an exponential model, extended to the multistage version in \cite{Shukla:2017,Shukla:2018}.


Concerning \ac{ev} dynamic load modelling,
the existing literature is limited.
The works in \cite{Islam:2010, Saha:2013, Dharmakeerthi:2014} include early attempts to
perform stability analyses of networks with \acp{ev} by analytically developing
a small-signal model. However, analyses are restricted to simple grids comprising
one generator and an \ac{ev} described by models
of different degrees of detail. A dynamic model was developed in \cite{Tian:2021} by simulating voltage disturbances at the \acp{ev} point of connection and fitting their power transient with a first-order response.
However, as the authors of \cite{Tian:2021} acknowledge, this model
does not always adequately represent \acp{ev}: indeed, based on the control parameters
of the converters in the \acp{ev} and \acfp{cs}, the order of the dynamic response may be
higher than one.

In this work, we propose accurate and computationally efficient static and dynamic load models of \acp{ev} connected to fast \acp{cs}, which can replace the standard representation as constant \textsc{pq} loads to assess more precisely the impact of rising \ac{ev} penetration during power system studies. Specifically, the contributions of this article are the following.

\begin{itemize} [leftmargin=*]
    \item A novel static load model is developed that mirrors the dependence of \ac{ev} power exchange on \ac{soc}, voltage, charging method, and the cathode chemistry of the battery pack. This
	last aspect is novel since previous works
	usually adopted Li-ion batteries
	without explicitly stating their cathode chemistry. In this work, we used some of the most popular ones
	\cite{Lebedeva:2018,Camargos:2022}: \ac{lfp}, \ac{lmo}, \ac{nmc}, and \ac{nca}.
	\item A novel dynamic load model based on \acl{vf} is proposed to describe the dynamic response of \acp{ev} during voltage disturbances. The shape of this response mirrors the controls installed in the converters of \acp{ev} and their \acp{cs}.
	\item Simulation results during static and dynamic grid studies of the \textsc{ieee14} system modified by adding loads given by \ac{ev} fleets are reported to validate the proposed models and highlight the inaccuracy introduced in some cases by the \ac{ev} representation as a constant \textsc{pq} load.
\end{itemize}


\section{Overview of static and dynamic load models}
\label{sec:load_model}
The next subsections outline some load models that are helpful for better understanding those proposed in this work. In general, a load model describes the relationship between the
voltage and power at the bus where the load is connected. For static and dynamic
load models, this relationship is respectively given by algebraic and differential equations
\cite{Ntovom:2022}.


\subsection{Static load models}
The most popular static load representations are the polynomial (\textsc{zip}) and exponential (\textsc{exp}) \cite{Kuronovic:2018} models, which relate load active power exchange $P$ to voltage magnitude $v$ as
\setlength{\arraycolsep}{0.0em}
\begin{eqnarray}
		P_\textsc{zip} &{}={}& P_\mathrm{nom} \left(k_0 + k_1 \left( \frac{v}{v_\mathrm{nom}} \right) + k_2 \left( \frac{v}{v_\mathrm{nom}} \right)^2 \right) \label{eq:static_zip} \\
		P_\textsc{exp} &{}={}& P_\mathrm{nom} \left( a_p + b_p \left( \frac{v}{v_\mathrm{nom}} \right)^{n_p} \right), \label{eq:static_exp}
\end{eqnarray}
\setlength{\arraycolsep}{5pt}
where $P_\mathrm{nom}$ is the load active power at nominal voltage $v_\mathrm{nom}$ (analogous formulas hold for the reactive power $Q$). Based on the values of $k_0$, $k_1$, and $k_2$, the \textsc{zip} model can synthesize constant power/current/impedance loads (or a mixture of those). As opposed to \textsc{zip} representation, the \textsc{exp} model can correctly reproduce also loads like air-conditioning systems \cite{Kurita:1988}, which are potentially detrimental to grid stability as their power absorption is inversely proportional to $v$ (i.e., ${n_p <0}$).
In these cases, using \textsc{zip} models could lead to inaccurate power system analyses. Other representations improve the above ones by adding terms that mirror the power dependence on frequency \cite{Kontis:2022} and resorting to multistage models \cite{Schneider:2011}.

\subsection{Dynamic load models}
\begin{figure}[!t]
	\centering
	\includegraphics[scale=1]{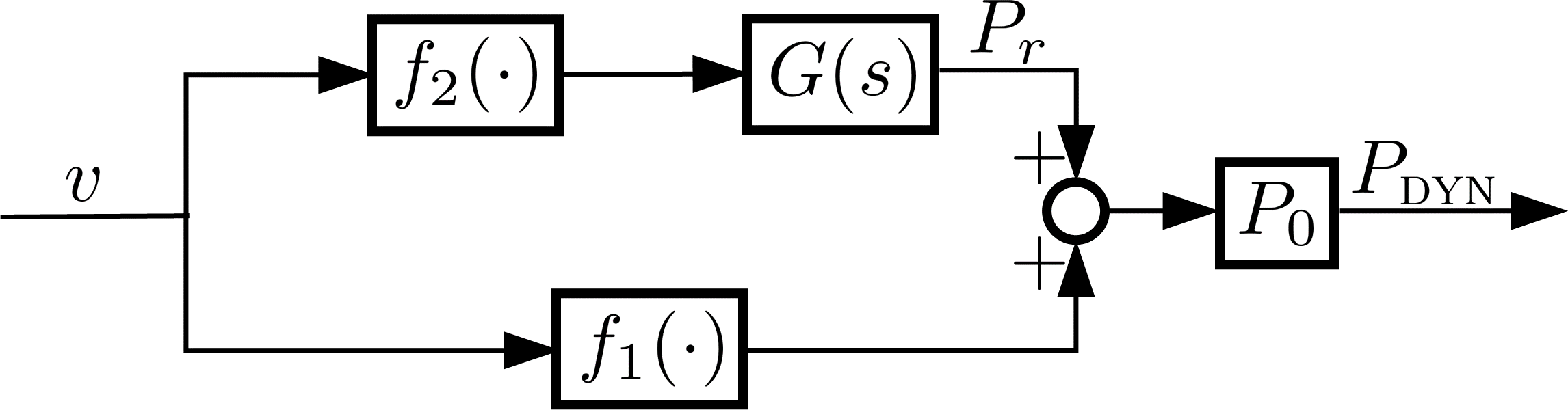}
	\caption{General block structure representation of the dynamic load model.}
	\label{F:dyn_load_model}
\end{figure}

Dynamic load models extend static representations and allow describing load behaviour during voltage disturbances. The models outlined hereafter can be implemented through the schematic in \RefFig{F:dyn_load_model}. The active power exchange is
\begin{equation}
	\label{eq:dyn_model1}
		P_\textsc{dyn}  = P_0 \left(f_1(v) + P_r \right) = P_0 \left(f_1(v) + G(s) f_2(v) \right),
\end{equation}
where 
$P_0$ refers to the power exchange before any disturbance occurs (an analogous formula holds for the reactive power). The $P_r$ term is the power recovery part and mirrors the power transient undergone by the load immediately after a voltage disturbance and before reaching a new steady-state value. The functions $f_1(v)$, $f_2(v)$ and the transfer function $G(s)$ give shape to this transient and their expression depends on the specific dynamic model used. In the literature, several formulations were developed, such as the \acf{erlm}, \acf{oclm}, and \acf{vflm}.

The \ac{erlm} describes a first-order response and is one of the oldest dynamic load representations. It was originally introduced to describe how \acp{oltc} connected to loads influenced dynamic power system behaviour \cite{Hill:1993}. In principle, the \ac{erlm} can be readapted to model the dynamics of other loads possibly lacking \acp{oltc} but falls short when higher-order responses need to be mimicked. For instance, this is the case of power electronics interfaced loads, whose $P_r$ component includes an exponential and a damped oscillatory part. To address this issue, one can resort to the \ac{oclm} and \ac{vflm} \cite{Kontis:2018,Paidi:2020}. The former models a second-order response, while the latter can describe responses of any order through \ac{vf}---an algorithm that approximates a measured or calculated frequency domain response with a rational transfer function $G(s)$\footnote{The interested reader is referred to \cite{Gustavsen:1999,Gustavsen:2006,Deschrijver:2008} for details about this algorithm and to \cite{VF:2016} for a downloadable \ac{vf} software package.}.

In the specific case of the \ac{vflm}, the expressions of $f_1(v)$, $f_2(v)$, and $G(s)$ are the following
\setlength{\arraycolsep}{0.0em}
\begin{eqnarray}
f_1(v)         &{}={}& \left(\frac{v}{v_0} \right)^{N_t} \nonumber \\
f_2(v)         &{}={}& \left(\frac{v}{v_0} \right)^{N_s} -  \left(\frac{v}{v_0} \right)^{N_t} \label{eq:dyn_model2} \\
G(s)           &{}={}& \sum\limits_{n=1}^{N_p} \frac{c_n}{s-a_n} \nonumber \,
\end{eqnarray}
\setlength{\arraycolsep}{5pt}
where $v_0$ is the load voltage before any disturbance occurs, and $N_p$ is the number of poles used with \ac{vf} to describe $G(s)$, whose poles and residues are $a_n$ and $c_n$, respectively. The work in \cite{Kontis:2018} explains the procedure required to derive $G(s)$ and the unknown parameters of this model (i.e., $N_t$, $N_s$, $c_n$, and $a_n$), which is not repeated here for the sake of brevity.

\section{Electric vehicle and charging station model}
\label{sec:ev_model}
The development of static and dynamic \ac{ev} load models requires executing three preliminary steps: (i) the identification of the \ac{ev} and \ac{cs} components that need to be modelled, (ii) the definition of the control scheme used to regulate \ac{ev} charging, and (iii) the selection of adequate topology and control parameters that reflect realistic \acp{ev} and \acp{cs}. Each of these phases is addressed in the next subsections.

\subsection{\ac{ev} and \ac{cs} topology}
\begin{figure}[!b]
  \centering
  \includegraphics[width=1\columnwidth]{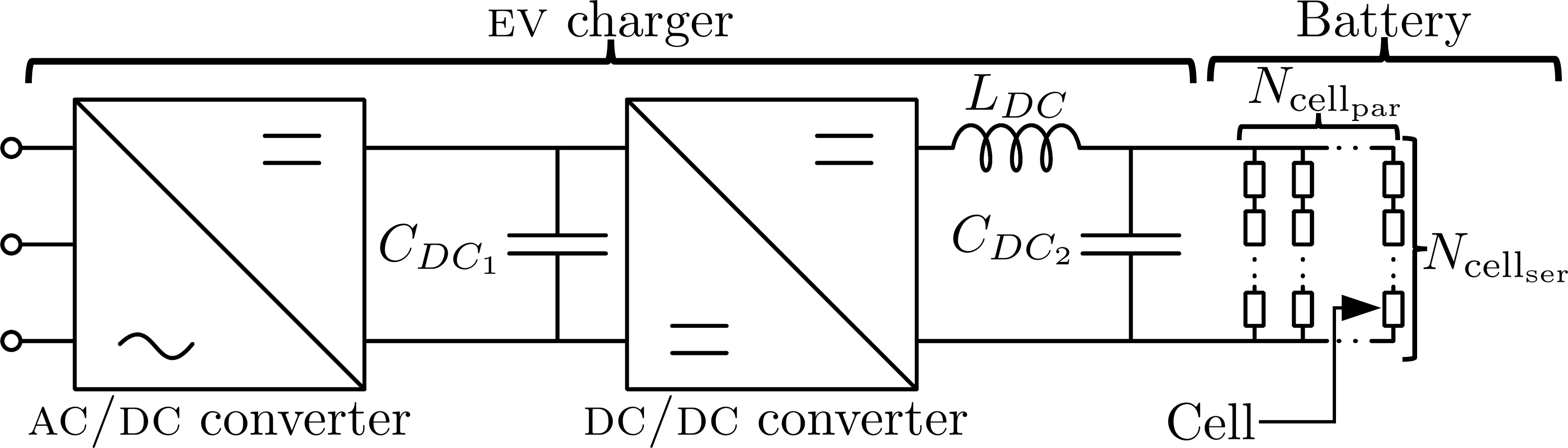}
  \caption{Synthetic schematic of the main components of an \ac{ev} and its \ac{cs}.}
  \label{F:charging_station}
\end{figure}

When parked for charging, the most relevant elements of the \ac{ev} are the battery pack and the charger, while the traction drive, motor, and other mechanical parts are neglected. Based on the application, the charger is either on-board or off-board the \ac{ev}. They are respectively suitable for a slow charge (e.g., through a household power outlet) and fast charge (i.e., through a dedicated \ac{cs}). This work focuses on fast \acp{cs}. \InRefFig{F:charging_station} depicts an implementation example: the charger consists of a two-stage \textsc{ac-dc/dc-dc} power conversion \cite{Haidar:2016}, while the battery pack comprises ${N_\mathrm{cell_{par}}}$ parallel branches of ${N_\mathrm{cell_{ser}}}$ series-connected identical cells \cite{Plett2:2015}.

\begin{figure}[!b]
	\centering
	\includegraphics[width=0.9\columnwidth]{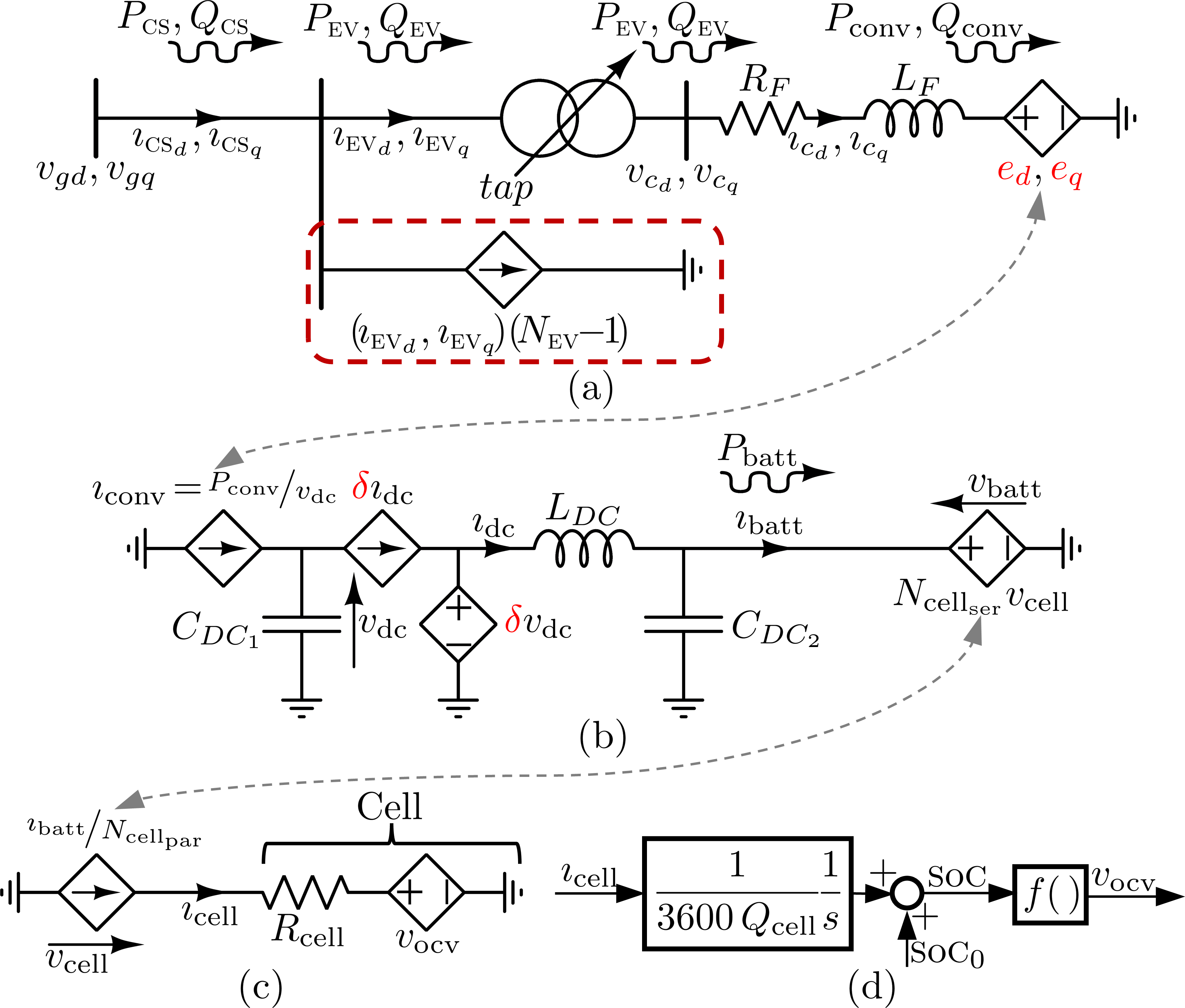}
	\caption{Schematic of the \ac{cs} and \ac{ev} at \textsc{ac} side (a) and \textsc{dc} side (b), single cell in a battery pack (c), the $\imath_\mathrm{cell}$-\textsc{ocv} relationship of a single cell (d). Dashed double arrows denote coupling between the dependent sources.}
	\label{F:topology}
\end{figure}

The different blocks of \RefFig{F:topology} depict the average model of the \ac{cs} in the \textsc{ac} and \textsc{dc} sides (including the cells in the \ac{ev} battery pack) adopted in this work. The variables highlighted in red are the output of the \ac{ev} charging control architecture described in the following, while those associated with a wavy arrow denote the active or reactive power at a given point.

\InRefFig{F:topology}(a) describes the \textsc{ac} side of the \textsc{ac/dc} converter through a single-phase schematic in the \textsc{dq} frame (ignore for now the circuit in the dashed section, whose purpose is explained in section \ref{sec:results}). It includes an $R_F$-$L_F$ filter and an $e_d$-$e_q$ dependent voltage source. An ideal transformer with adjustable tap ratio is added to the \ac{cs} to serve a twofold purpose. First, it facilitates the connection of \acp{cs} to any bus of the benchmark system used in the following (which may operate at different nominal voltages) by selecting a proper transformation ratio. Second, by changing its tap ratio during power flow and transient simulations, it allows investigating \ac{ev} behaviour and deriving static and dynamic load models---an approach commonly used also for other loads \cite{Kontis:2018}. 

As shown in \InRefFig{F:topology}(b), the \textsc{dc} side of the \textsc{ac/dc} converter is given by a dependent current source that acts as an ideal power coupler between the two sides of the converter. The two dependent sources $\delta \imath_\mathrm{dc}$ and $\delta v_\mathrm{dc}$ (together with $C_{DC_1}$, $C_{DC_2}$, and  $L_{DC}$) implement the average model of the \textsc{dc/dc} converter described in \cite{Canizares:2016}. Lastly, a dependent voltage source mirrors the voltage given by the series connection of $N_\mathrm{cell_{ser}}$ cells, which are assumed to behave in the same way.


\InRefFig{F:topology}(c) depicts the model of a single cell. The current dependent source relates the cell and battery currents (i.e., $\imath_\mathrm{cell}$ and $\imath_\mathrm{batt}$) through the number of parallel branches $N_\mathrm{cell_{par}}$. The single cell can be described through models of different degrees of accuracy \cite{Plett1:2015}. In this work, the cell comprises an equivalent series resistance $R_\mathrm{cell}$ and a dependent voltage source which imposes the \ac{ocv}.
In turn, as shown in \RefFig{F:topology}(d), this voltage depends on the cell's \ac{soc} and current through a non-linear relationship $f(\,)$.
In \cite{Tremblay:2009,Fernandez:2018}, this relationship is approximated by
\begin{equation}
	\label{eq:ocv_analytical}
	f  = v_{\mathrm{ocv}} (\ac{soc}_0)  =   E_0 - K \frac{1 - \ac{soc}_0}{\ac{soc}_0} Q + A e^{(1-\ac{soc}_0) Q} ,
\end{equation}
where $Q$ and $\ac{soc}_0$ are respectively the capacity and initial \ac{soc} of the whole battery pack, while $E_0$, $K$, and $A$ are specific cell parameters that vary with the chemistry of the cell.

\begin{figure}[!b]
	\centering
	\includegraphics[width=1\columnwidth]{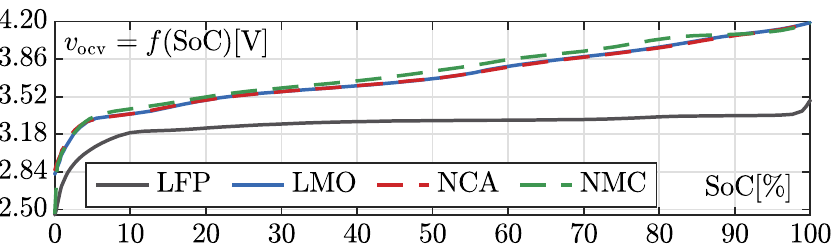}
	\caption{\ac{ocv}-\ac{soc} relationship of the Li-ion cell chemistries considered \cite{Tran:2021}.
	}
	\label{F:soc_curves}
\end{figure}

In this work, rather than resorting to analytical expressions, we used the \ac{ocv}-\ac{soc} relationships of the Li-ion cathode chemistries cited in the introduction and shown in \RefFig{F:soc_curves}, which have been obtained by interpolating the experimental results reported in Fig.~6 of \cite{Tran:2021}.

\subsection{\ac{cs} control architecture}
\begin{figure}[!t]
	\centering
	\includegraphics[width=1\columnwidth]{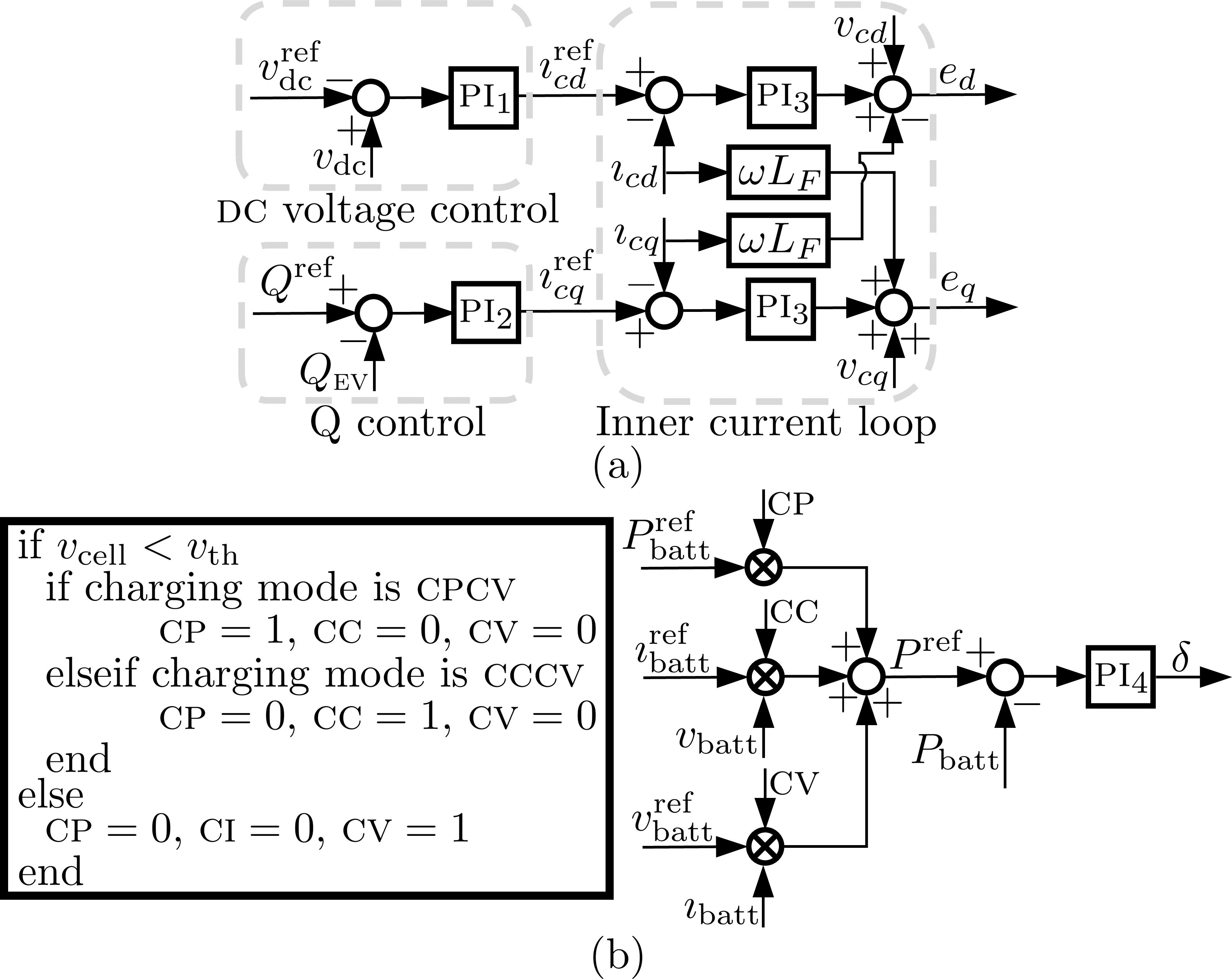}
	\caption{Control architecture of the \textsc{ac/dc} (a) and \textsc{dc/dc} (b) converters.}
	\label{F:control}
\end{figure}

The two converters in the \ac{cs} used in this work are entrusted with different tasks. As shown in \RefFig{F:control}(a), the \textsc{ac/dc} converter implements vector current control and exploits variables formulated in the \textsc{dq} frame \cite{Yazdani:2010}. It is in charge of keeping the voltage $v_\mathrm{dc}$ across the capacitor $C_{DC_1}$ close to the value $v_\mathrm{dc}^\mathrm{ref}$ and ensuring that the \ac{ev} exchanges no reactive power $Q_\textsc{ev}$ (i.e., power factor correction is implemented and, thus, $Q^\mathrm{ref} = 0$). These control loops define a reference value for the current $\imath_c$ (see \RefFig{F:topology}(a) for its location) on the direct and quadrature axis, which are translated by an inner current loop into the $e_d, e_q$ components of the dependent voltage source.

\InRefFig{F:control}(b) depicts the \textsc{dc/dc} converter control scheme. Its purpose is to regulate the duty cycle $\delta$ of the converter so that the power absorbed by the battery pack matches a reference value $P^\mathrm{ref}$. This value is defined according to the pseudo-code in \RefFig{F:control}(b), which controls the battery based on the charging mode employed (i.e., either \ac{cccv} or \ac{cpcv}) and the difference between the cell voltage $v_\mathrm{cell}$ and a given threshold $v_\mathrm{th}$ \cite{Plett2:2015}. During charge, one variable at a time among $P_\mathrm{batt}$, $\imath_\mathrm{batt}$, and $v_\mathrm{batt}$ (see \RefFig{F:topology}(b) for their locations) is regulated to follow a reference value denoted by the superscript $^\mathrm{ref}$. In turn, as shown in \RefFig{F:control}(b), these variables determine $P^\mathrm{ref}$.

\subsection{\ac{ev} and \ac{cs} specifics}
\label{S:EV_CS_DATA}
To accurately assess the impact of rising electric mobility on the grid, the synthesis of  equivalent static and dynamic load models must stem from a realistic design of \acp{ev} and \acp{cs}.

In this regard, in the simulations described in the following sections, we considered fast \acp{cs} of ${P_{\textsc{ev}_\mathrm{nom}} \! = \! 50 \, \kilo\watt}$ nominal power (a value compatible with \cite{Habib:2018}) and the 2017 Tesla Model~S as reference \ac{ev}, given by a ${E_{\textsc{ev}_\mathrm{nom}} \! = \! 75 \, \kilo\watt\hour}$ battery \cite{Sanguesa:2021} and an assumed nominal $\textsc{dc}$ voltage of ${V_{\textsc{ev}_\mathrm{nom}} \! = \! 400 \, \volt}$.

As to the cathode chemistry of the Li-ion battery pack (information rarely disclosed by \ac{ev} manufacturers), we selected those cited in the introduction, associated with the data in Table \ref{T:data} \cite{Tran:2021}. For each of them, the number of series-connected cells and parallel branches equals respectively ${N_\mathrm{cell_{ser}} \! = \! \text{round} \left( \frac{V_{\textsc{ev}_\mathrm{nom}}}{v_{\mathrm{cell}_\mathrm{nom}}} \right)}$ and ${N_\mathrm{cell_{par}} \! = \! \text{round} \left (\frac{E_{\textsc{ev}_\mathrm{nom}}}{V_{\textsc{ev}_\mathrm{nom}} Q_{\mathrm{cell}}} \right)}$, with $Q_\mathrm{cell}$ being the nominal cell capacity.

\begin{figure}[!t]
	\centering
	\includegraphics[width=1\columnwidth]{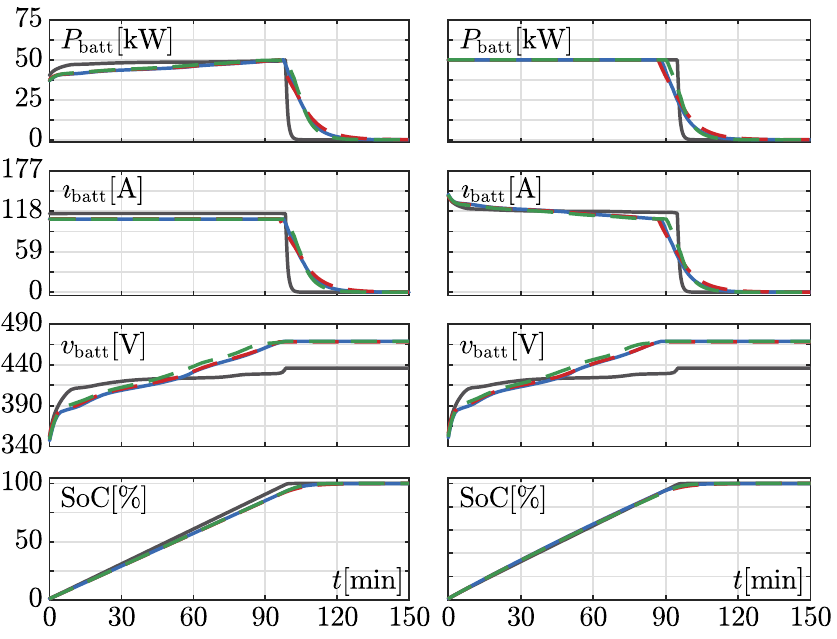}
	\caption{From top to bottom: evolution of $P_\mathrm{batt}$, $\imath_\mathrm{batt}$, $v_\mathrm{batt}$, and \ac{soc} for the benchmark \ac{ev} with \ac{cccv} (left panels) and \ac{cpcv} (right panels) charging modes and different Li-ion chemistries. Curves are colour-coded as in \RefFig{F:soc_curves}.}
	\label{F:cccv_cpcv}
\end{figure}

Lastly, the remaining parameters in Table \ref{T:data} were used to design the \ac{cs} and its control architecture. \InRefFig{F:cccv_cpcv} shows the evolution of some battery variables obtained with the abovementioned design during a full \ac{ev} charge (i.e., from $0$ to $100 \%$ \ac{soc})  for each chemistry and charging mode.

\begin{table}[!t]
	\caption{Electric vehicle and charging station data \label{T:data}}
	\vspace{-3mm}
	\centering
	\begin{tabular}{ccccc}
		\multicolumn{5}{c}{Battery cell parameters} \\ \hline
		\multicolumn{1}{c|}{Chemistry}     & \multicolumn{1}{c|}{$v_\mathrm{cell_{nom}} [\volt]$}   & \multicolumn{1}{c|}{$Q_\mathrm{cell} [\ampere\hour]$}     & \multicolumn{1}{c|}{$R_\mathrm{cell} [\ohm]$}      & \!\!\!\!\! $v_\mathrm{th} [\volt]$ \!\!\!\!\!     \\
		\multicolumn{1}{c|}{\ac{lfp}}            & \multicolumn{1}{c|}{3.20}          & \multicolumn{1}{c|}{2.6}              & \multicolumn{1}{c|}{0.053}              & \multicolumn{1}{c}{3.488}     \\
		\multicolumn{1}{c|}{\ac{lmo}}            & \multicolumn{1}{c|}{3.70}          & \multicolumn{1}{c|}{2.6}              & \multicolumn{1}{c|}{0.080}              & \multicolumn{1}{c}{4.188}     \\
		\multicolumn{1}{c|}{\ac{nca}}            & \multicolumn{1}{c|}{3.60}          & \multicolumn{1}{c|}{3.2}              & \multicolumn{1}{c|}{0.058}              & \multicolumn{1}{c}{4.188}     \\
		\multicolumn{1}{c|}{\ac{nmc}}            & \multicolumn{1}{c|}{3.60}          & \multicolumn{1}{c|}{2.0}              & \multicolumn{1}{c|}{0.080}              & \multicolumn{1}{c}{4.183}     \\ \hline
		\multicolumn{5}{c}{Charging station parameters} \\ \hline
		\multicolumn{1}{c|}{$R_F [\milli\ohm]$}     & \multicolumn{1}{c|}{$L_F [\milli\henry]$}   & \multicolumn{1}{c|}{$C_{DC_1} [\milli\farad]$}     & \multicolumn{1}{c|}{$L_{DC} [\milli\henry]$}      & \!\! $C_{DC_2} [\milli\farad]$ \!\!     \\
		\multicolumn{1}{c|}{3.20}            & \multicolumn{1}{c|}{0.2}          & \multicolumn{1}{c|}{1}              & \multicolumn{1}{c|}{0.2}              & \multicolumn{1}{c}{0.5}     \\ \hline
		\multicolumn{5}{c}{Control reference values} \\ \hline
		\multicolumn{1}{c|}{$v_\mathrm{dc}^\mathrm{ref} [\volt]$} & \multicolumn{1}{c|}{$Q^\mathrm{ref} [\textsc{var}]$} & \multicolumn{1}{c|}{$P_\mathrm{batt}^\mathrm{ref} [\kilo\watt]$} & \multicolumn{1}{c|}{$v_\mathrm{batt}^\mathrm{ref} [\volt]$} & $\imath_\mathrm{batt}^\mathrm{ref} [\ampere]$ \\[0.3em]
		\multicolumn{1}{c|}{800}            & \multicolumn{1}{c|}{0}          & \multicolumn{1}{c|}{50}              & \multicolumn{1}{c|}{\!\!\!\!\! $N_\mathrm{cell_{ser}} \! v_\mathrm{th}$ \!\!\!\!\!}              & \multicolumn{1}{c}{\!\!\! $\nicefrac{P_\mathrm{batt}^\mathrm{ref}}{v_\mathrm{batt}^\mathrm{ref}}$ \!\!\!}             \\ \hline
		\multicolumn{5}{c}{\textsc{pi} regulator parameters} \\ \hline
		\multicolumn{1}{c|}{}               & \multicolumn{1}{c|}{$\textsc{pi}_1$}          & \multicolumn{1}{c|}{$\textsc{pi}_2$}              & \multicolumn{1}{c|}{$\textsc{pi}_3$}              & $\textsc{pi}_4$              \\
		\multicolumn{1}{c|}{$k_p$}             & \multicolumn{1}{c|}{0.01}          & \multicolumn{1}{c|}{0}              & \multicolumn{1}{c|}{0.142}              & 0.001              \\
		\multicolumn{1}{c|}{$k_i$}             & \multicolumn{1}{c|}{1000}          & \multicolumn{1}{c|}{33}              & \multicolumn{1}{c|}{43.909}              & 1
	\end{tabular}
\end{table}

\section{Simulation results}
\label{sec:results}

\begin{figure}[t!!!]
	\begin{center}
		\includegraphics[width=1\columnwidth]{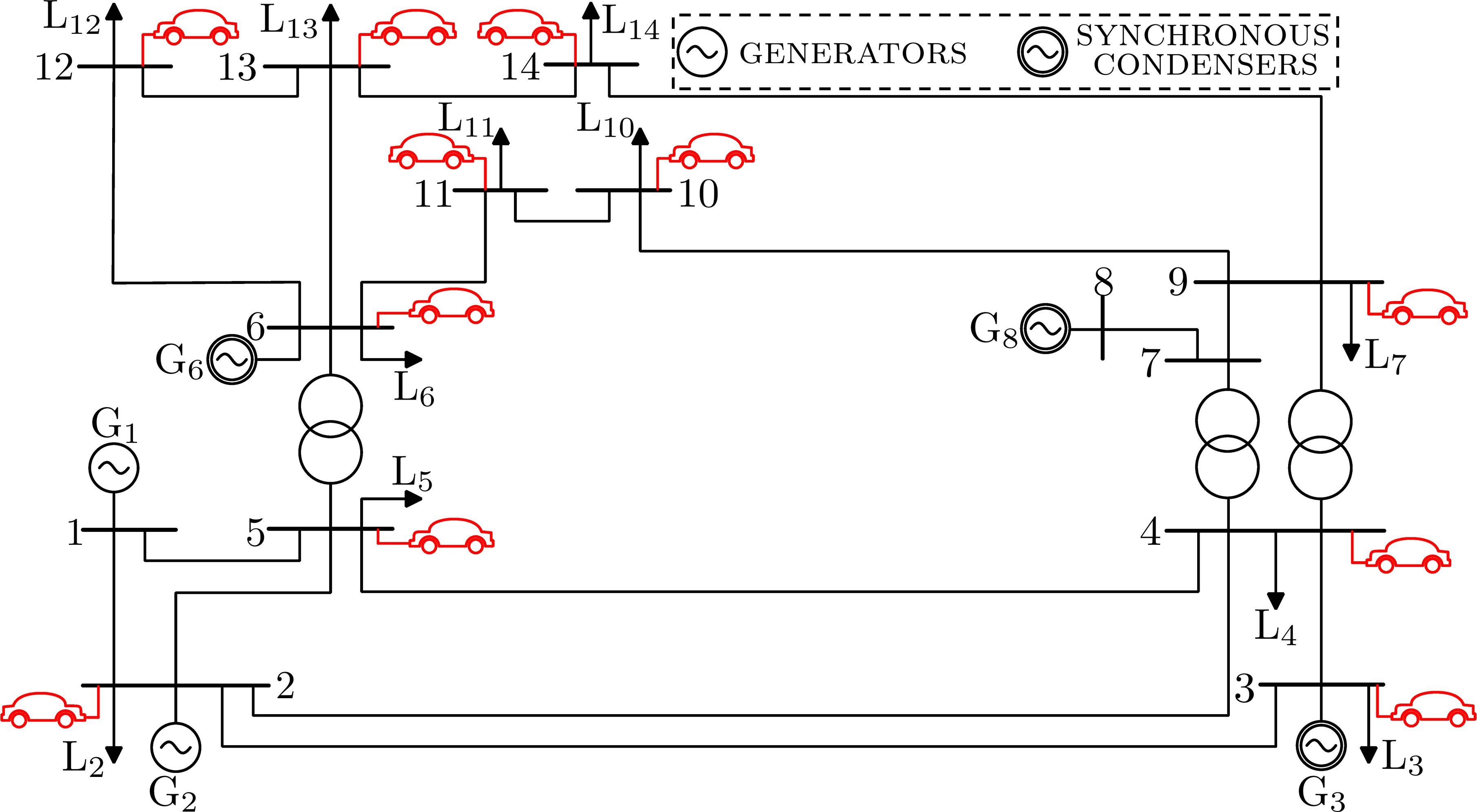}
	\end{center}
	\caption{The schematic of the modified \textsc{ieee14} power system with \ac{ev} fleets.}
	\label{F:ieee14}
\end{figure}

In this section, computationally efficient and accurate static and dynamic load models of \acp{ev} are developed by using as a benchmark the \textsc{ieee14} power system, whose parameters can be found in \cite{Milano:2010}\footnote{The simulation results shown hereafter were obtained with PAN simulator \cite{Bizzarri:2014,Bizzarri:2017,Linaro:2022}. The files needed to run the simulations are available on GitHub at \url{https://github.com/Davide-del-Giudice/Electric_vehicle_models.git}}. The obtained models are also used for static and dynamic grid studies involving fleets of \acp{ev}, as shown in \RefFig{F:ieee14}. Such analyses prove that the proposed models allow gathering results comparable to those obtained with detailed \ac{ev} representations and highlight the inaccuracy introduced in some cases by modelling \acp{ev} as simple constant \textsc{pq} loads.

\subsection{Static \ac{ev} load model derivation}
\label{S:static_single}
We first simulated the charge of a single \ac{ev} through a fast \ac{cs} in different cases. In each case, we considered one of the previously mentioned cell chemistries, charging modes, and a set of initial $\ac{soc}$ values  (i.e., $\ac{soc}_0$) ranging from 10 to $90\%$. By acting on the tap of the transformer in \RefFig{F:topology}(a), we varied the \textsc{ac}-side voltage magnitude at the point of connection of the \textsc{ac/dc} converter (hereafter referred to as $v_c$) around its rated value ${v_{c_\mathrm{nom}} \!\! = \! \sqrt{v_{c_d}^2 \! + \! v_{c_q}^2} \! = \! 230 \, \volt}$ and recorded the \ac{ev} active power normalized with respect to ${P_{\textsc{ev}_\mathrm{nom}} \!\! = \! 50 \, \kilo\watt}$.

\begin{figure}[!t]
	\centering
	\includegraphics[width=1\columnwidth]{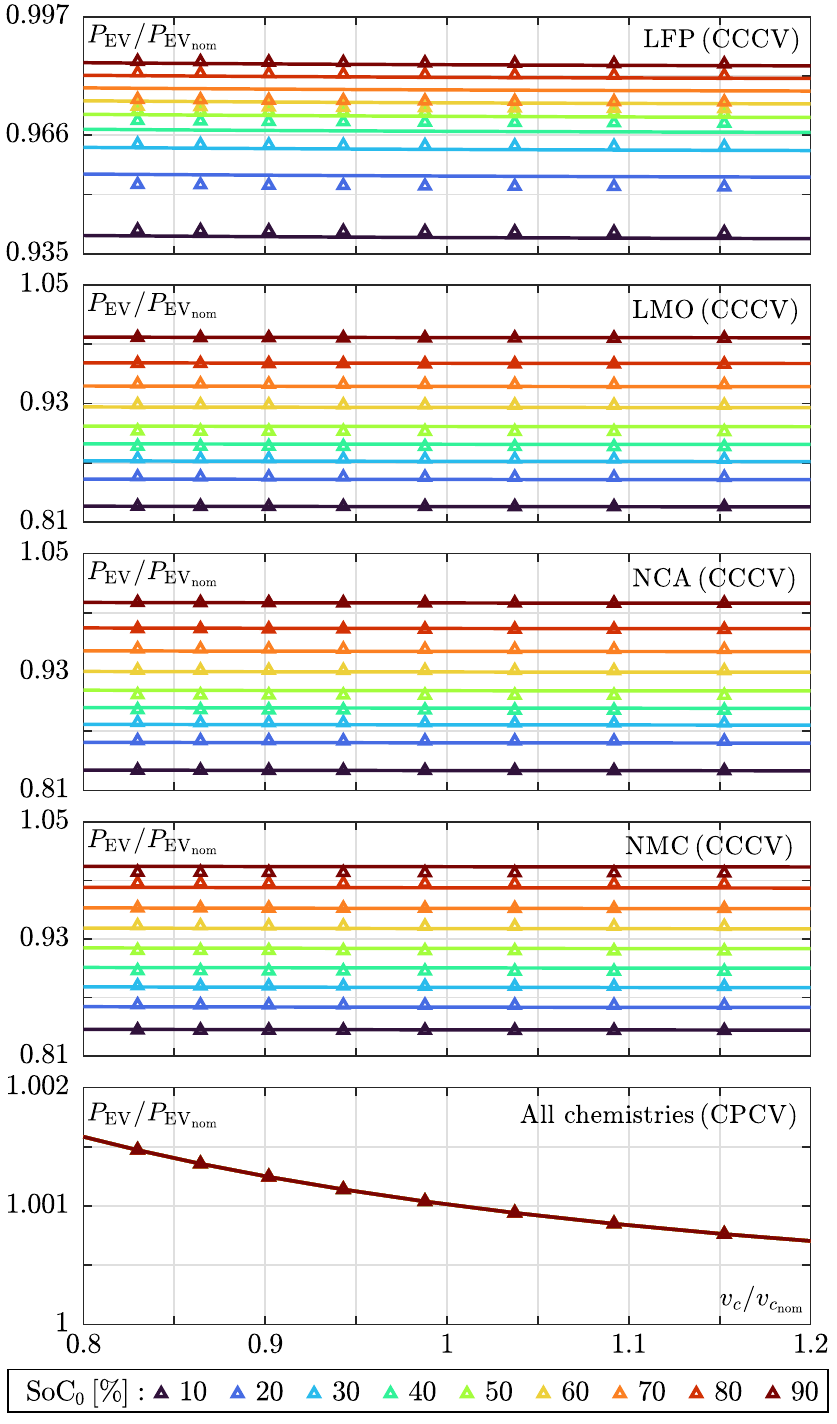}
	\caption{Voltage-dependent active power $P_\textsc{ev}$ (normalized w.r.t. ${P_{\textsc{ev}_\mathrm{nom}}}$) for different cathode chemistries, charging modes, and $\ac{soc}_0$ values (each associated with a colour in the legend). The triangular markers correspond to the results obtained through power flow simulations and using the complete \ac{ev} and \ac{cs} model of section \ref{sec:ev_model}. The solid lines are the results obtained by fitting the previous curves with \eqref{eq:static_ev_exp} and \eqref{eq:static_exp} for \ac{cccv} and \ac{cpcv} charging, respectively. Notice the reduced interval of variation of $P_{\textsc{ev}}$ when the \ac{lfp} chemistry is employed compared to that obtained with other cathode chemistries when \ac{cccv} charging is adopted.}
	\label{F:static_cccv_cpcv}
\end{figure}

So doing, the results in \RefFig{F:static_cccv_cpcv} were obtained (neglect for now the solid lines). Analogous plots for the reactive power are not shown since power factor correction is implemented (thus, ${Q_{\textsc{ev}} \! = \! 0}$). For each panel, which corresponds to a specific cathode chemistry and charging method, the triangular coloured markers  denote the \ac{ev} power-voltage relationship for a given value of $\ac{soc}_0$ specified in the legend.

By looking at the plots, the following points can be raised.
\begin{itemize} [leftmargin=*]
	\item With \ac{cccv} charging, the active power exchange $P_\textsc{ev}$ increases with $\ac{soc}_0$. The magnitude of this trend depends on the cathode chemistry considered. Results show that \ac{lfp} cells have a maximum power variation (i.e., the relative difference between the markers at $\ac{soc}_0 = 10\%$ and $\ac{soc}_0 =90\%$) of about $5 \%$\footnote{One can notice a similar power variation in Fig.2(b) of \cite{Shukla:2018}. Therefore, in that paper we assume an \ac{lfp} chemistry was considered for the battery pack, even though the authors do not explicitly state it.}. This value is lower than that of other cathode chemistries ($\approx 20 \%$) that behave alike due to the similarity of their \ac{ocv}-\ac{soc} curves in \RefFig{F:soc_curves}.
	\item On the contrary, with \ac{cpcv} charging, $P_\textsc{ev}$ does not change with $\ac{soc}_0$ (i.e.,  all coloured traces are superposed on each other so that only one is visible). Moreover, the results obtained do not vary with the cathode chemistry. The only visible trend is that $P_\textsc{ev}$ slightly decreases as the voltage $v_c$ increases. This feature was discussed in \cite{Saha:2014,Tian:2021} and is mainly due to the power losses of the \ac{ev} charger filter.
\end{itemize}

Based on the above, in the case of \ac{cpcv} charging, one could fit the active power exchange $P_\textsc{ev}$ of the \ac{ev} and its \ac{cs} through the \textsc{exp} model of \eqref{eq:static_exp}, which only accounts for voltage dependency.  On the contrary, this expression is inadequate when considering \ac{cccv} charging due to the previously mentioned dependence on $\textsc{soc}_0$. To address this issue, some works in the literature resort to a multistage model. For instance, in \cite{Shukla:2017,Shukla:2018} \RefEq{eq:static_exp} is still used to fit power exchange $P_\textsc{ev}$ for different values of $\textsc{soc}_0$, thereby obtaining look-up tables.

In this work, instead of performing multiple fittings for each $\textsc{soc}_0$ value, we adopted a novel and compact static model that can be intuitively derived as follows. From \RefFig{F:topology}(a) and power conservation, the \ac{ev} active power exchange $P_\textsc{ev}$ is the sum of that of the \textsc{ac/dc} converter $P_\mathrm{conv}$ and the losses of the filter resistance $R_F$. In turn, assuming lossless \textsc{ac/dc} and \textsc{dc/dc} converters, by looking at \RefFig{F:topology}(b) one can derive that $P_\mathrm{conv}$ equals the battery power  $P_\mathrm{batt}$. In case of constant current charging,  ${P_\mathrm{batt} \! = \! v_\mathrm{batt} \imath_\mathrm{batt} \! = \! N_\mathrm{cell_{ser}} v_\mathrm{cell} \imath_\mathrm{batt}^\mathrm{ref}}$. Since $N_\mathrm{cell_{ser}}$ and $\imath_\mathrm{batt}^\mathrm{ref}$ are known parameters (see Table \ref{T:data}), the only unknown left to be determined is $v_\mathrm{cell}$. From \RefFig{F:topology}(c), neglecting for simplicity the voltage drop across the cell resistance $R_\mathrm{cell}$, we have ${v_\mathrm{cell} \! \approx \! v_\mathrm{ocv}}$. Therefore, ${P_\mathrm{conv} \! \approx \! N_\mathrm{cell_{ser}} v_\mathrm{ocv} \imath_\mathrm{batt}^\mathrm{ref}}$.

Based on the above, for \ac{cccv} charging, we fitted $P_\textsc{ev}$ as
\begin{equation}
	\label{eq:static_ev_exp}\!\!\!
		P_\textsc{ev} \!\! = \!\! P_{\textsc{ev}_\mathrm{nom}} \!\!\! \left(\!\! \underbrace{b_p \!\! \left(\!\! \frac{v_c}{v_{c_\mathrm{nom}}}\!\! \right)^{\!\!\! n_p}\!}_{\text{V dependence}} \!\!\! + \! \underbrace{c_p \!\! - \!\! d_p \frac{1 \!\! - \!\! \ac{soc}_0}{\ac{soc}_0} \!\! + \!\! e_p e^{\!-\! f_p(\! 1 -   \ac{soc}_0 \!)}\!}_{\text{$\ac{soc}_0$ dependence}} \!\! \right),
\end{equation}
where $a_p$, $b_p$, $n_p$, $c_p$, $d_p$, $e_p$, and $f_p$ are the unknown parameters of the model. This expression includes two parts, which respectively mirror the power exchange dependence on voltage and $\ac{soc}_0$. The former is primarily due to the power losses of the filter resistance $R_F$, which are such that $P_\textsc{ev}$ decreases as voltage increases (i.e., ${n_p \! < \! 0}$) \cite{Saha:2014}. On the contrary, the latter is due to $P_\mathrm{conv}$. As already stated and shown in \RefFig{F:soc_curves}, $v_\mathrm{ocv}$ is a function of \ac{soc}, which can be approximately described with \RefEq{eq:ocv_analytical} regardless of cathode chemistry. This suggests that similar terms can approximate $P_\mathrm{conv}$ due to its dependence on $v_\mathrm{ocv}$. Indeed, $\ac{soc}_0$ dependence in \eqref{eq:static_ev_exp} retraces \RefEq{eq:ocv_analytical}.

The {\tt fit} command in MATLAB was used to fit the curves in each panel of \RefFig{F:static_cccv_cpcv} with \RefEq{eq:static_ev_exp} and \RefEq{eq:static_exp} for the cases based on \ac{cccv} and \ac{cpcv} charging, respectively. Table \ref{T:static} reports, for each chemistry and charging mode, the corresponding fitted parameters and \ac{rmse} as an indicator of the goodness-of-fit. The solid lines in \RefFig{F:static_cccv_cpcv}, colour-coded as the triangular markers, correspond to the fitted $P_\textsc{ev}/P_{\textsc{ev}_\mathrm{nom}}$ results obtained in the same range of $v_c/v_{c_\mathrm{nom}}$ and $\textsc{soc}_0$ values. The comparison between the markers and lines, as well as the \ac{rmse} values in Table \ref{T:static}, prove the adequacy of the fitting.

Both the static models of \RefEq{eq:static_exp} and \RefEq{eq:static_ev_exp} can be easily implemented in any simulator. Moreover, they lead to a computational burden comparable to that obtained by traditionally modelling \ac{ev} as constant \textsc{pq} loads but, as shown in the next subsection, are more accurate.

\begin{table}[!b]	
	\caption{Fitted parameters of the proposed \ac{ev} static load model and \textsc{rmse} values \label{T:static} (in some cells, \textsc{n.a.} stands for "not available").}
	\vspace{-2mm}
	\centering
	\setlength\tabcolsep{3.5 pt}
	\begin{tabular}{c|c|c|c|c|c|c|c|c}
		\hline
		Chemistry                                                            & $a_p$ & $b_p$ & $n_p$ & $c_p$ & $d_p$ & $e_p$ & $f_p$ & \textsc{rmse}            \\ \hline
		\ac{lfp} (\ac{cccv})                                                           & \textsc{n.a.}    & 0.001 & -2    & 0.90 & 0.003 & 0.09 & 0.34 & 0.002           \\ \hline
		\ac{lmo} (\ac{cccv})                                                           & \textsc{n.a.}    & 0.001 & -1.8 & 0.75 & 0.003 & 0.27 & 1.11 & 0.003           \\ \hline
		\ac{nca} (\ac{cccv})                                                           & \textsc{n.a.}    & 0.001 & -2.9 & 0.77 & 0.003 & 0.26 & 1.17 & 0.002           \\ \hline
		\ac{nmc} (\ac{cccv})                                                           & \textsc{n.a.}    & 0.001 & -2    & 0.07 & 0.001 & 0.96 & 0.23 & 0.003           \\ \hline
		\begin{tabular}[c]{@{}c@{}} All chemistries \\ (\ac{cpcv})\end{tabular} & 1     & 0.001 & -2    & \textsc{n.a.}    & \textsc{n.a.}    & \textsc{n.a.}    & \textsc{na}    & $< \! 10^{-6}$ \\ \hline
	\end{tabular}
\end{table}

\subsection{Static power system studies with fleets of \acp{ev}}
\label{S:static_fleet}
After analysing the active power dependency of one \ac{ev} on different aspects (i.e., voltage, \ac{soc}, charging method, and cathode chemistry) and deriving a static load model, we performed static power system studies by overloading the \textsc{ieee14} network through the connection of \ac{ev} fleets in parallel to the original loads (see \RefFig{F:ieee14}). In particular, we considered three cases where \acp{ev} and fast \acp{cs} are modelled differently.

In the first case (base case), \ac{ev} fleets are represented through constant \textsc{pq} loads
as in most literature works.
These loads do not withdraw reactive power (i.e., \acp{cs} implement active power correction), while their active power amounts to $\lambda P_{\mathrm{nom}_i}$. $P_{\mathrm{nom}_i}$  is the
nominal power of a given load L$_i$ (with ${i\in\{1,2,...,14\}}$) to which an \ac{ev} fleet is connected in parallel, while $\lambda$
is the overload percentage due to \ac{ev} penetration, which is assumed to be $20 \%$.  To balance generation,
the active power provided by the generators considered as \textsc{pv} busses in power flow
simulations was multiplied by $\lambda$, too. Unmatched power demand is covered by the slack
bus (i.e., generator G$_1$).

In the second case, \acp{ev} and \acp{cs} are described  with the detailed model in Figs. \ref{F:charging_station}--\ref{F:control}
and employ the already mentioned different cathode cell chemistries and charging modes.
In these cases, every new load added to those already existing
in the benchmark consists of a fast \ac{cs} to which ${N_{\textsc{ev}_i}}$ \acp{ev} are
connected in parallel. In particular,
${N_{\textsc{ev}_i} = \text{round} \left(\nicefrac{\lambda P_{\mathrm{nom}_i} }{P_{\textsc{ev}_\mathrm{nom}}} \right)}$, where $P_{\textsc{ev}_\mathrm{nom}}$ is
the nominal power of a single \ac{cs} (i.e., $50 \, \kilo\watt$ as stated in section
\ref{S:EV_CS_DATA}). The parallel connection of \acp{ev} in each \ac{cs} has been synthesized
through the dependent current source in the dashed section of
\RefFig{F:topology}(a)\footnote{This synthesis holds if
\acp{ev} in a \ac{cs} behave in the same way: this assumption suffices for the
scenarios shown in this work, which only aim at validating static
and dynamic \ac{ev} and \ac{cs} load models.
A more realistic approach requires retaining the behaviour of each \ac{ev} as
their \ac{soc}, chemistry, charging mode, and even location could differ.
However, this calls for (i) the
exploitation of static and dynamic models like those presented here
instead of detailed representations to limit computational burden and (ii) the addition of the
probabilistic approaches mentioned in the introduction that "modulate" the $N_\textsc{ev}$ term of this paper (which are outside the scope of this work).}.  So doing, we assume the contemporary charge of $N_{\textsc{ev}_i}$ \acp{ev} per \ac{cs}.

Lastly, in the third case, \acp{ev} and \ac{cs} are replaced with
the static \ac{ev} load models of \RefEq{eq:static_exp} and \RefEq{eq:static_ev_exp} using the parameters in Table \ref{T:static} and adequately scaling the nominal \ac{ev} power $P_{\textsc{ev}_\mathrm{nom}}$ to attain the same overload level $\lambda$.

\begin{figure}[!t]
	\centering
	\includegraphics[width=1\columnwidth]{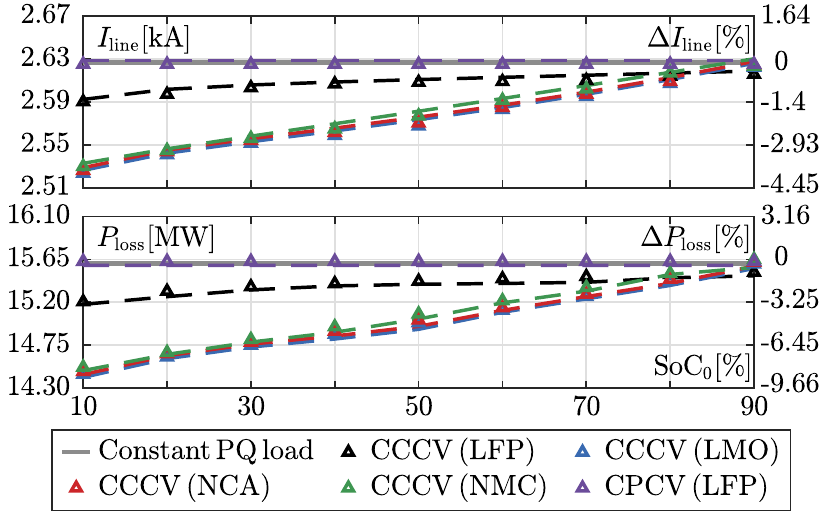}
	\caption{Simulation results of the \textsc{ieee14} system obtained with ${\lambda = 20 \%}$ and in the cases listed in the legend. When considering \ac{cpcv} charging, only the case with \ac{lfp} chemistry is reported since the other ones basically yield identical results. The results denoted by triangular markers were obtained by using the detailed model of \acp{ev} and \acp{cs} in Figs. \ref{F:charging_station}--\ref{F:control}. The solid gray line represents the results obtained by replacing the \ac{ev} fleets with constant \textsc{pq} loads (base case), while the dashed lines (colour-coded as the markers) represent those obtained with the static models of \RefEq{eq:static_ev_exp} and \RefEq{eq:static_exp}. The panels depict the line current between \textsc{bus1} and \textsc{bus2} (top) and the total active power losses (bottom) as a function of $\ac{soc}_0$ in physical values (left y-axis) and in relative percentage w.r.t. the base case (right y-axis).
	}
	\label{F:overload_static}
\end{figure}

Some of the results obtained by simulating the above scenarios and considering different values of $\ac{soc}_0$ are reported in \RefFig{F:overload_static} (refer to the caption for the meaning of the markers and dashed lines). The top panel shows the current flowing through the line that connects \textsc{bus1} and \textsc{bus2} (analogous trends hold for the currents in the other lines), while the bottom panel shows the total active power losses of the system.

By comparing each case, one can notice that the base one leads to the worst-case scenario (i.e., highest currents and losses). Very similar results are obtained with \ac{cpcv} charge. On the contrary, with \ac{cccv} charge, lower currents and losses are obtained as smaller values of $\textsc{soc}_0$ are considered. For example, when $\ac{soc}_0 = 10 \%$, the relative percentage variation in line current (power) with respect to the base case is close to $-1.6 \%$ ($-2.9 \%$) with \ac{lfp} cathode chemistry, while with the other chemistries it is about ${-4.2 \%}$ ($-7.6 \%$).

These results confirm that \acp{ev} and \acp{cs} can be adequately replaced in static power system studies by constant \textsc{pq} loads when \ac{cpcv} charging is used. On the contrary, with \ac{cccv} charging, this substitution becomes less acceptable as \ac{ev} penetration increases and $\ac{soc}_0$ decreases. If so, adopting the static model in \RefEq{eq:static_ev_exp} is necessary to retain accuracy.

\subsection{Dynamic \ac{ev} load model derivation}
\label{S:dynamic_single}
We now derive an \ac{ev} dynamic load model based on \ac{vf}. As in section \ref{S:static_single}, we considered again a single \ac{ev} charged through a fast \ac{cs}. To examine its dynamic behaviour during a disturbance and derive a \ac{vflm}, we simulated a $-3\%$ voltage decrease by varying the tap of the ideal transformer in \RefFig{F:topology}(a). For space reasons, we discuss only the results obtained by assuming ${\ac{soc}_0=10\%}$ and adopting \ac{cpcv} charging and the \ac{lfp} cathode chemistry (analogous results hold for the other cases).

\begin{figure}[!t]
	\centering
	\includegraphics[width=1\columnwidth]{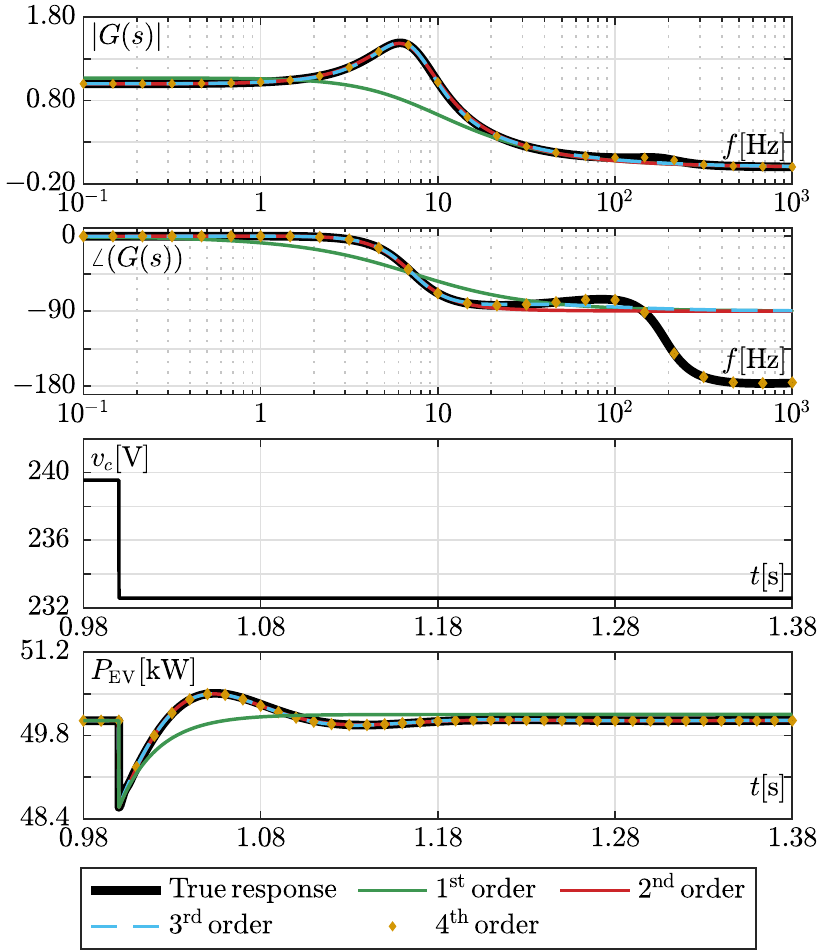}
	\caption{First and second panel from the top: magnitude and phase, in degrees, of $G(s)$ for the detailed \ac{ev} and \ac{cs} model and for \acp{vflm} of order ranging from one to four. Third and fourth panel: voltage step and corresponding \ac{ev} active power response obtained in the real case and with the \acp{vflm}.
		\label{F:single_dynamic}}
\end{figure}

\InRefFig{F:single_dynamic} depicts the results obtained: the first two panels from the top show the magnitude and phase of $G(s)$ (derived from the detailed \ac{ev} model of section \ref{sec:ev_model}) and the relative fits of different orders derived with \ac{vf}. While the first-order model is inadequate in the frequency range considered, those of order from two to three deviate from the true behaviour of $G(s)$ for frequencies higher than $10\, \hertz$. On the contrary, the fourth-order \ac{vflm} always gives an accurate fit.

The other panels of \RefFig{F:single_dynamic} show the voltage disturbance simulated and the resulting \ac{ev} active power response in the real case and for the \acp{vflm} of different order. Other than confirming the inadequacy of the first-order model, the bottom panel highlights that increasing the \ac{vflm} order from two to four does not substantially change the accuracy of the fitted active power response. Thus, if dynamics of frequencies higher than $10 \, \hertz$ are not of interest, this suggests that a second-order \ac{vflm} is enough to describe the dynamic \ac{ev} behaviour.

Lastly, it is worth noting that the \ac{vflm} needs as a prerequisite accurate pre-disturbance values $P_0$ and $v_0$ (see \RefFig{F:dyn_load_model} and \RefEq{eq:dyn_model2}) \cite{Milano:2010}. As it is done here, this can be accomplished by pairing the proposed \ac{vflm} with the previously discussed static model in \RefEq{eq:static_ev_exp} or \RefEq{eq:static_exp}, based on the charging mode used. During power flow analyses, the static model allows deriving the correct \ac{ev} steady-state operating condition and, thus, $P_0$ and $v_0$. These values are used as input for the \ac{vflm} during small-signal or transient analyses.

\subsection{Dynamic power system studies with fleets of \acp{ev}}
\label{S:dynamic_fleet}
Lastly, we determined the maximum amount of \acp{ev} that can be connected to the \textsc{ieee14} benchmark before instability occurs.  The authors of \cite{Moschella:2020} had a similar aim but described \acp{ev} with constant \textsc{pq} loads. As proved in section \ref{S:static_fleet}, this simple model adequately represents \acp{ev} only in \ac{cpcv} charging mode during static grid studies. However, in the case of dynamic grid studies and depending on the \ac{ev} and \ac{cs} converter control parameters, replacing \acp{ev} with constant \textsc{pq} loads may lead to wrong results even if \acp{cpcv} charging is employed. To ensure accuracy also in these cases, dynamic load models like those described in this work are needed.

To validate these statements, we extended the work in section \ref{S:static_fleet} and swept the parameter $\lambda$ across a given interval, assuming ${\ac{soc}_0 \!=\! 10\, \%}$. For each value of $\lambda$, we performed a power flow and a pole-zero analysis (i.e., stability analysis)\footnote{The adoption of average \ac{ev} and \ac{cs} models formulated in the \textsc{dq} frame is essential to perform these analyses. Indeed, it leads to the absence of harmonics, and both negative and zero sequence components. Only under these conditions, the network steady-state operation in the \textsc{abc} frame corresponds to an equilibrium point in the \textsc{dq} frame \cite{Bizzarri:2020}. During pole-zero analyses, the grid model is linearized around this constant solution, thus obtaining a linear time-invariant system, whose attainment of poles and zeros is commonplace.
}. So doing, one can derive the minimum value of $\lambda$ such that the real part of at least one eigenvalue of the system switches from negative to positive (i.e., instability occurs).
For brevity, we discuss only the results obtained with \ac{cpcv} charging and the \ac{lfp} cathode chemistry. As in  section \ref{S:static_fleet}, we considered three cases, each corresponding to a different  \ac{ev} and fast \ac{cs} model. In the first case, we adopted the accurate model of Figs. \ref{F:charging_station}--\ref{F:control}.  In the second case, constant \textsc{pq} loads replace the fleet of \acp{ev}. Lastly, in the third case we adopted the second-order \ac{vflm} of section \ref{S:dynamic_single}.
Moreover, we analysed how results change when the parameter $k_i$ of the regulator $\textsc{pi}_1$ equals 1000, 100, and 80 (in each case, the \ac{vflm} parameters were updated to ensure a proper fit). The reason for this was twofold: first, to exemplify the impact of the tuning of converter controls in \acp{ev} and \acp{cs} on grid stability; second, to highlight the accuracy of the proposed dynamic model.


\begin{figure}[!t]
	\centering
	\includegraphics[width=1\columnwidth]{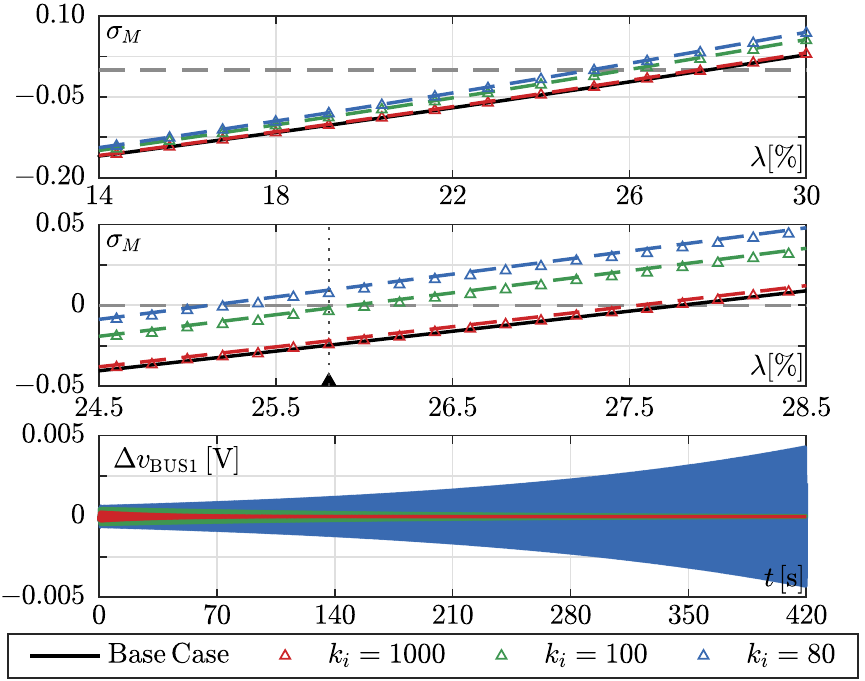}
	\caption{Top panel: maximum real part $\sigma_M$ of the eigenvalues of the modified \textsc{ieee14} system
		obtained by sweeping $\lambda$ from $14\%$ to $30\%$ and considering the $k_i$ values of the $\textsc{pi}_1$ regulator shown in the legend. The gray dashed line is placed at $\sigma_M = 0$. The triangular markers show the results obtained with the detailed \ac{ev} and \ac{cs} model in Figs. \ref{F:charging_station}--\ref{F:control}. The solid black line represents the results obtained by replacing \ac{ev} fleets with constant \textsc{pq} loads, while the dashed lines represent those achieved with the second-order \ac{vflm} of section \ref{S:dynamic_single}. The middle panel is an inset of the top panel. Bottom panel: voltage deviations at \textsc{bus1} around its steady-state value due to a load disturbance, obtained for different values of $k_i$ and ${\lambda \! = \! 25.8 \%}$.
		\label{F:overload_dynamic}}
\end{figure}
The top and middle panel of \RefFig{F:overload_dynamic} show how the maximum real part of the eigenvalues of the \textsc{ieee14} benchmark evolve in a given range of $\lambda$ for each case (refer to the caption for the meaning of the markers and the solid and dashed lines). These panels highlight two aspects. First, based on the control parameters in \RefFig{F:control}, constant \textsc{pq} loads may not adequately reflect the true behaviour of \acp{ev} and \acp{cs}, and not even lead to the worst-case scenario. Indeed, as $k_i$ decreases, the pole-zero analysis results obtained by replacing \ac{ev} fleets with \textsc{pq} loads increasingly differ from the true ones\footnote{If \ac{cccv} charging were considered, instability would occur for values of $\lambda$ higher than those obtained with \ac{cpcv} charging and shown in \RefFig{F:overload_dynamic}. This happens because, for the same value of $\lambda$, the actual level of stress in the grid  (e.g., higher currents and grid losses) due to \ac{ev} fleets is generally lower with \ac{cccv} than \ac{cpcv} charging, regardless of $\ac{soc}_0$. This feature, which can be observed in \RefFig{F:overload_static}, translates into higher stability margins.}. Second, the very good match between the triangular markers and dashed lines of the same colour proves that the proposed second-order \ac{vflm} can correctly replace detailed \ac{ev} and fast \ac{cs} models in stability studies, regardless of the tuning of their converter control parameters.

As a further proof of the results obtained with the pole-zero analyses, we performed transient simulations of the modified \textsc{ieee14} benchmark by assuming $\lambda \!=\! 25.8 \%$ (black arrowhead in the middle panel of \RefFig{F:overload_dynamic}) and considering the previous values of $k_i$. Based on the pole-zero analysis and the above value of $\lambda$, the case with $k_i = 80$ is unstable, whereas those with ${k_i \! = \! 100}$ and $1000$ lead to a stable system (indeed, for $\lambda \!=\! 25.8 \%$, only in the first case there is at least one pole with positive real part). This is confirmed by the bottom panel of \RefFig{F:overload_dynamic}, which shows the voltage deviations at \textsc{bus 1} around its steady-state value after a small load disturbance.
As expected, only when ${k_i \! = \! 80}$ are voltage oscillations not damped: in fact, they diverge rather slowly because for ${\lambda \! =\! 25.8 \%}$ the positive real part of the unstable pole is close to zero\footnote{It is also worth pointing out that the voltage oscillations obtained in the unstable case for $k_i \! = \! 80$ have a frequency lower than $10 \, \hertz$, which justifies the adequacy of adopting a second-order \ac{vflm}. If that were not the case, based on what was stated in section \ref{S:dynamic_single}, a higher order should be used.}.


\section{Conclusions}
\label{sec:conclusions}
We have presented accurate and computationally efficient static and dynamic load models for \acp{ev} and fast \acp{cs}, which take into account the dependence of active power exchange on voltage, \ac{soc}, charging method, cathode chemistry, and converter controls. These models can replace the standard \ac{ev} representation as constant \textsc{pq} loads to assess more reliably the impacts of electric mobility during static and dynamic grid studies. Through the simulation results of the \textsc{ieee14} benchmark, we have validated the accuracy of the proposed models and highlighted that in some cases the standard \ac{ev} representation as constant \textsc{pq} loads lacks accuracy.

In particular,  we have shown that, when \ac{cccv} charging is employed, the battery cathode chemistry and \ac{soc} significantly affect \ac{ev} power exchange. This behaviour cannot be adequately mimicked through constant \textsc{pq} loads.  Moreover, due to the \ac{ev} and \ac{cs} control system parameters, this simplified representation may not adequately reflect the dynamic behaviour of \acp{ev}, even if \ac{cpcv} charging is used.


\color{black}

\end{document}